# Implementing strong interference in ultrathin film top absorbers for tandem solar cells

Yifat Piekner[1], Hen Dotan[2], Anton Tsyganok[2], Kirtiman Deo Malviya[2], Daniel A. Grave[2], Ofer Kfir[3,4] and Avner Rothschild[2*]

[1] The Nancy & Stephen Grand Technion Energy Program (GTEP), Technion – Israel Institute of Technology, Haifa 32000, Israel

[2] Department of Materials Science and Engineering, Technion – Israel Institute of Technology, Haifa 32000, Israel

[3] Solid State Institute and Physics Department, Technion–Israel Institute of Technology, Haifa 32000, Israel.

[4] 4th Physical Institute, University of Göttingen, Göttingen 37077, Germany.

*avner@mt.technion.ac.il

**Abstract**

Strong interference in ultrathin film semiconductor absorbers on metallic back reflectors has been shown to enhance the light harvesting efficiency of solar cell materials. However, metallic back reflectors are not suitable for tandem cell configurations because photons cannot be transmitted through the device. Here, we introduce a method to implement strong interference in ultrathin film top absorbers in a tandem cell configuration through use of distributed Bragg reflectors (DBRs). We showcase this by designing and fabricating a photoelectrochemical-photovoltaic (PEC-PV) stacked tandem cell in a V-shaped configuration where short wavelength photons are reflected back to the photoanode material (hematite, $\alpha\text{-Fe}_2\text{O}_3$), whereas long wavelength photons are transmitted to the bottom silicon PV cell. We employ optical simulations to determine the optimal thicknesses of the DBR layers and the V-shape angle to maximize light absorption in the ultrathin (~10 nm thick) hematite film. The DBR spectral response can be tailored to allow for a more than threefold enhancement in absorbed photons compared to a layer of the same thickness on transparent current collectors. Using a DBR to couple a bottom silicon PV cell with an ultrathin hematite top PEC cell, we demonstrate unassisted solar water splitting and show that DBRs can be designed to enhance strong interference in ultrathin films while enabling stacked tandem cell configuration.

**Introduction**

Tandem solar cells are often used to surpass single junction efficiencies and harness a larger range of the solar spectrum.[1] For some applications such as photoelectrochemical (PEC) water splitting, tandem cell configurations are necessary when additional photovoltage is required to drive the electrochemical reaction.[2] To efficiently utilize the solar spectrum, the large band gap absorber is placed on top of the small band gap absorber in tandem solar devices. Introducing wavelength selective mirrors, such as distributed Bragg reflectors (DBRs), between the absorbers can be used to optimize the efficiency by reflecting back photons with high energy to the top absorber while transmitting photons with low energy to the bottom absorber.[3]

When using thin-film solar cells, efficiency may be improved by increasing the absorption within the thin film.[4–8] Employing strong interference in ultra-thin semiconducting absorbing layers using specular metallic back reflectors can significantly enhance broadband light absorption.[8–10] However, metallic back-reflectors are not suited for use in stacked tandem solar cell devices[11] since they do not transmit photons to the bottom cell. In this work, we show that the specular metallic back reflectors can be replaced with optimized DBRs to yield strong interference within the ultra-thin top absorber while efficiently transmitting light to the bottom absorber for stacked tandem cell operation.

To demonstrate this method, we combine detailed materials fabrication and characterization with optical simulations to design a PEC-PV tandem cell for hydrogen production via solar water splitting for direct conversion of solar energy into storable chemical fuel.[12,13] We optimize DBRs to maximize high energy photon absorption in a top ultra-thin film hematite ($\alpha$-$Fe_2O_3$) photoanode while transmitting low energy photons to a bottom Si PV module. We use hematite ($\alpha$-$Fe_2O_3$) due to its visible light absorption,[14] stability in alkaline solutions,[15] vast abundance, nontoxicity and low cost. Moreover, hematite has a suitable bandgap energy for tandem configuration with Si photovoltaic (PV) cells.[1,16] PEC-PV tandem cells with hematite-based PEC cells in front of Si PV modules can generate sufficient photovoltage for unassisted solar water splitting.[17] However, hematite possesses poor charge transport properties.[18–21] Due to its small charge collection length, we use thin top hematite films to enable the photo-generated carriers to travel a short distance to the photoanode/electrolyte interface and thereby improve the charge extraction. Using thin thicknesses, however, limits the light absorption, and hence, the light harvesting efficiencies. Overcoming this tradeoff is possible by combining strong interference with photon re-trapping in V-shaped structures using specular metallic (e.g., silver) back-reflectors at the cost of wasted absorption in the metal.[22] It is noteworthy that besides the re-trapping effect, the V-shaped structure also modifies the incidence angle of the incoming light, thereby increasing the effective travel length of the light within the photoactive layer. In principle, this trivial geometric effect could also be utilized in planar devices, without the V-shaped structure. However, in this case the reflected light is lost, unlike the V-shaped structure that re-traps the reflected light that bounces back and forth between two mirrors facing each other at an angle.[23] Since the reflection increases drastically at high incidence angles (see Figure S1), this effect alone has limited potential to enhance light harvesting in planar solar cells with ultra-thin films. In this work, we combine both wave optics (interference) and ray optics (multiple reflections) to enhance light harvesting in ultra-thin films. We show that replacing the metallic back-reflector reported in our previous work[22] by a DBR, which is a semi-transparent dielectric mirror, enables simple construction of stacked PEC-PV tandem cells that can be tailored to enhance strong interference in an ultra-thin absorber while reducing wasted absorption.

DBRs are widely used in optoelectronics, especially as thin film filters.[24] They are composed of stratified planar structures with alternating layers of two distinct dielectric materials with different refractive indices. The inward propagating transmitted beam interferes with the reflected beams from the interfaces between the alternating layers. The interference can be tailored to transmit a selected range of wavelengths through the DBR stack whereas the other wavelengths are back-reflected. In our case, the DBRs are tailored to reflect the higher-energy photons ($\lambda < 590$ nm) of the incident solar radiation back into the upper hematite photoanode and transmit the lower-energy photons into the bottom Si PV cell. The reflection intensity depends on the difference between the refractive indices of the dielectric materials in the DBR stack[25,26] and the number of bilayer pairs.[26–29] The spectral response is determined by the thickness of the dielectric layers.[3,28,29] The large difference in the refractive indices of $Nb_2O_5$ and $SiO_2$, dielectric materials that do not compete with hematite and Si for absorption of above bandgap photons, make them well-suited for our DBRs. Another advantage of the $SiO_2$-$Nb_2O_5$ DBRs, besides enabling stacked tandem cell configuration, is that they replace the silver back-reflectors in our previous design[22] with cheaper materials that are more compatible with the device fabrication process as well as with operation in alkaline solutions. Tuning the DBR stack so that the multiple reflected beams constructively interfere in such a way to enable the strong interference effects is more challenging than in the case of a single metallic back reflector where light is only reflected at one interface. Specifically, high surface quality and precise thicknesses are instrumental to assure the multiple reflections are plane-wave-like, with uniform constructive interference within the active layer.

SiO$_2$-TiO$_2$ DBRs were employed recently to couple BiVO$_4$ / WO$_3$ porous photoanodes with dye solar cells to construct PEC-PV tandem cells that achieved a solar-to-hydrogen conversion efficiency of 7.1%.[30] Unlike those BiVO$_4$ / WO$_3$ photoanodes that do not benefit from the strong interference effects due to their mesoporous structure,[30] here we use compact ultrathin hematite films that give rise to these effects, as reported in our previous study.[22] Therefore, the DBR optimization in this work is fundamentally different than in reference 29[30], as it accounts for the strong interference effects and photon re-trapping in V-shaped cell configurations.[22] We show that by employing the strong interference effects combined with photon re-trapping in a V-shaped cell configuration, the light harvesting yield can be more than tripled within ultra-thin (10 nm) hematite layers.

Besides the DBR stack, a transparent electrode is needed to collect the current from the photoanode. Indium tin oxide (ITO) thin films are best suited for this purpose[31] due to their remarkable combination of high electrical conductivity (~16 Ω/□)[32] and high optical transparency (~80%)[33] in the wavelength range for productive absorption in both the hematite photoanode and the Si PV cell. However, these two properties have opposite dependences on the film thickness: increasing the ITO thickness increases its conductivity but decreases its transparency.[34] These considerations should also be taken into account in the design of the PEC-PV tandem cell.

We show that optimal coupling of our hematite-Si PEC-PV cells using SiO$_2$-Nb$_2$O$_5$ DBRs as wavelength-selective dielectric mirrors and V-shaped structures significantly improves photoelectrochemical performance. Our samples display a remarkable enhancement of 101% in the plateau photocurrent for 10 nm thick hematite photoanodes coated with ultrathin Fe$_{1-x}$Ni$_x$OOH co-catalyst overlayer[35] using DBRs with tailored spectral response and a 90º V-shaped structure. At the reversible potential of water oxidation, 1.23 V$_{RHE}$, the photocurrent is enhanced by a factor of four with respect to the control hematite photoanode of the same thickness without the DBR stack and without the co-catalyst.

To achieve optimal performance, the thicknesses of all the layers in the stack are carefully optimized in order to maximize absorption in the ultrathin hematite film as well as the transmission of sub-bandgap photons ($h\nu < 2.1$ eV) to the bottom Si PV cell. The optical optimization was carried out by calculating the optical response of the stack using a customized optical modeling algorithm. A code employing this algorithm to simulate the spectral response of complex optical stacks at any incident illumination angle has been included in the supporting information. The optical parameters of all the individual layers were measured by spectroscopic ellipsometry. Prior to the optical optimization of the DBR stack, we first found the optimal thickness of the hematite film and the $SnO_2$ underlayer so as to maximize the absorbed photon to current efficiency (APCE) for water photo-oxidation. The next step was finding the optimal thickness of the ITO current collector in order to reach maximal absorption in the hematite film. Subsequently, we optimized the DBR stack so as to maximize the incident photon to current efficiency (IPCE) of the entire stack. We then added an ultrathin $Fe_{1-x}Ni_xOOH$ co-catalyst overlayer to reduce the onset potential. For additional photocurrent enhancement we introduced the V-shaped cell configuration to re-trap back-reflected photons that escaped out of the hematite photoanode. Finally, we demonstrated unassisted solar water splitting system by coupling the hematite-based PEC cell with a Si PV mini-module. The results are presented in the order of the optimization sequence.

**Results and discussion**

**Hematite, $SnO_2$ underlayer and ITO film thickness optimization**

The first optimization step is empirical optimization of the film thickness of the hematite layer and $SnO_2$ underlayer to reach maximal APCE. We use Sn-doped (1 cation%) hematite which was identified as an optimal donor dopant in a previous study.[36] Based on previous work on the beneficial role of ultrathin underlayers to enhance the photocurrent of hematite photoanodes,[37–39] we employed undoped $SnO_2$ underlayers to enhance the performance of our ultrathin hematite photoanodes (see Figure S2).[22,40] The $SnO_2$ underlayer ability to prevent holes generated in the hematite from reaching the substrate is commonly attributed to the large valence-band offset between $SnO_2$ and hematite.[14] Furthermore, unlike many of the other candidate materials, $SnO_2$ is stable in alkaline solutions, as can be assessed by looking at the Pourbaix diagram of tin.[41] Both the hematite and SnO$_2$ films were deposited on top of ITO current-collector layers. All layers were fabricated by pulsed laser deposition (PLD) using similar process recipes to those described elsewhere.[40]

A series of samples with different thickness of hematite and SnO$_2$ layers was fabricated. Their optical absorption spectra and PEC water splitting performance were measured to find the optimal thicknesses that yield the maximal APCE for water photo-oxidation at the photocurrent plateau region. The photocurrent – potential voltammograms and absorptance (A= 1 - transmittance - reflectance) measurements are presented in Figure S2 and Figure S3, respectively. The color map in Figure 1(a) shows the plateau photocurrent (measured under front solar-simulated illumination at a potential of 1.6 V vs. the reversible hydrogen electrode, $V_{RHE}$) as a function of the hematite and SnO$_2$ layer thicknesses. The color map in Figure 1(b) shows the respective APCE values obtained by dividing the plateau photocurrent values by the absorbed photocurrent, $J_{abs} = q \times \varphi_{abs}$, where $\varphi_{abs}$ is the photon flux absorbed within the hematite layer (obtained by calculations) and q is the electron charge. The APCE reaches a maximum for ~10 nm (±1 nm) thick hematite layer and ~2 nm for the SnO$_2$ underlayer. Therefore, these layer thicknesses were selected for the subsequent optimization steps.

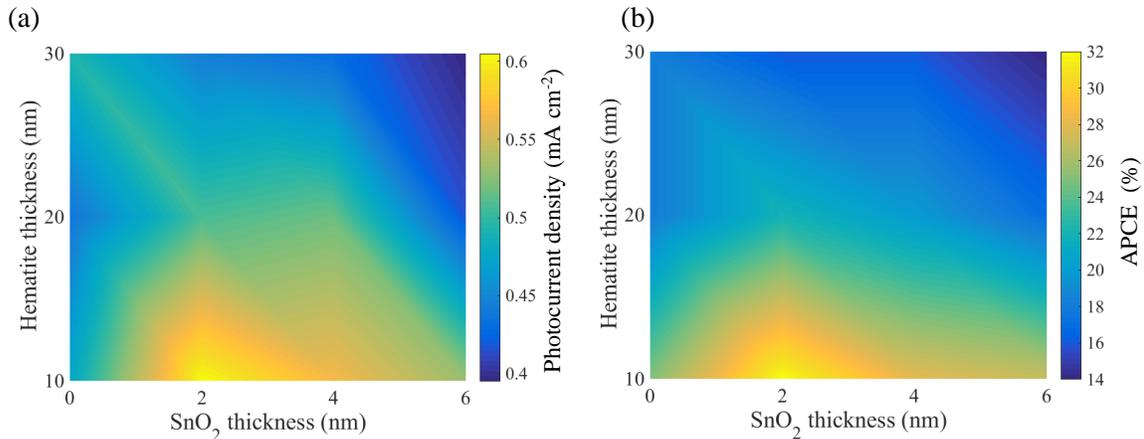

**Figure 1**: (a) The plateau photocurrent (measured at 1.6 $V_{RHE}$) and (b) APCE values as a function of the hematite and SnO$_2$ layer thickness.

The next step was to find the optimal thickness of the ITO layer in order to achieve both low resistance and high transparency (in the 350-1100 nm spectral range). The latter is important to minimize parasitic absorption within the ITO layer in order to maximize the intensity of the back-reflected light that goes back into the hematite layer. Toward this end, we deposited hematite and $SnO_2$ underlayer stacks with the previously found optimal thicknesses (10 and 2 nm, respectively) on top of ITO layers with different thickness. The ITO layers were deposited by PLD on glass substrates (see Experimental Section). The photocurrent – potential voltammograms (measured under front solar-simulated illumination) of three photoanodes with ITO thickness of 82, 162 and 242 nm (determined by spectroscopic ellipsometry) are presented in Figure 2(a). Unexpectedly, the photoanode with the smallest ITO film thickness (82 nm, green curve) reached the highest photocurrent, followed by the one with the thickest ITO layer (242 nm, blue curve), and last the one with the intermediate ITO thickness (162 nm, red curve). The dependence of the photocurrent on the ITO film thickness cannot be ascribed to trivial electrical effects because the energy band alignment is expected to be the same for all three photoanodes. Furthermore, the photoanode with the thinnest ITO layer and highest series sheet resistance displays the highest photocurrent. Therefore, this must be an optical effect that arises from interference between forward and backward propagating waves from the incident and back-reflected beams, respectively. This hypothesis is confirmed by optical calculations of the amount of photons absorbed within the 10 nm thick hematite film as a function of the ITO film thickness, presented in Figure S4. The optical calculations show a peak for an ITO thickness of ~100 nm, followed by a dip at 150 nm, and then another smaller peak at 230 nm. The wavy pattern arises from interference effects.[24] The photocurrent values measured for the three photoanodes at a potential of 1.6 $V_{RHE}$ are overlaid on the calculated dashed-line curve in Figure S4, showing a reasonably fair agreement between the photoelectrochemical measurements and the optical calculations.

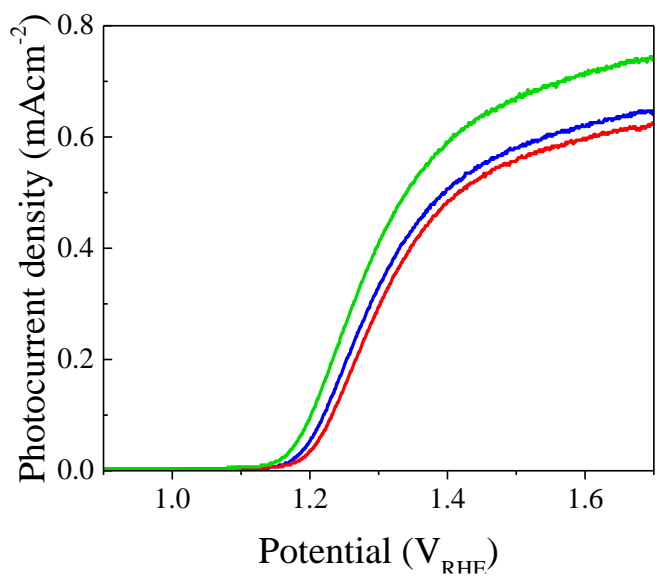

**Figure 2:** Photocurrent voltammograms of photoanodes comprised of 10 nm thick hematite layer, 2 nm thick $SnO_2$ underlayer and ITO layer with thickness of 82 nm (green), 162 nm (red) and 242 nm (blue) on top of eagle glass substrates.

According to our optical calculations presented in Figure S4, the optimal ITO thickness is ~100 nm. It is noteworthy that once more layers are added to the stack (e.g. with the DBR stack in place) the optimal thickness may change. Hence, the absorption in hematite may be improved by iterative thickness optimization after determining the DBR layer thickness. The sheet resistance values of the ITO layers are presented in Table 1, and their transmittance and reflectance spectra are presented in Figure S5. The rest of the light is assumed to be absorbed in the ITO layer giving rise to parasitic absorption (optical loss) in the non-photoactive layers. The diffuse reflectance is negligible due to the very smooth surface. The arithmetical mean deviation of the surface roughness, $R_a$, of these specimens is 0.3 nm, measured by AFM (see Figure S6). The calculated[24,42] total integrated scatter from a surface with 0.3 nm root-mean-square (RMS) surface roughness is ~0.01%. This value is almost three orders of magnitude lower than the value calculated for commercial transparent current collectors with RMS surface roughness of 8 nm[43] (see SI for total integrated scatter calculation). As expected, the optical loss due to parasitic absorption in the ITO layer increases with increasing ITO thickness. The cross-section TEM micrograph in Figure S7 shows a high quality smooth ITO layer obtained by direct deposition on top of smooth eagle glass substrate. The obtained ITO layer smoothness, transparency and conductivity enable it to serve as an optimal current collector in simple stacked PEC-PV configuration.

**Table 1:** The plateau photocurrent (at 1.6 $V_{RHE}$), ITO sheet resistance and parasitic absorption of the photoanodes presented in Figure 2.

|            | Photocurrents at 1.6 $V_{RHE}$ [mA/cm$^2$] | ITO sheet resistance [Ω/□] | Parasitic absorption [mA/cm$^2$] |
|---|---|---|---|
| **240 nm ITO** | 0.62 ± 0.01 | 41.9 ± 0.1 | 0.57 |
| **160 nm ITO** | 0.60 ± 0.01 | 70.3 ± 0.3 | 0.34 |
| **82 nm ITO**  | 0.72 ± 0.01 | 74.1 ± 0.4 | 0.29 |

**DBR stack optimization**

After optimizing the ITO/SnO$_2$/hematite photoanode stack to yield maximal APCE, our next step was to design a DBR stack that maximizes the light harvesting within the 10 nm thick hematite film of the chosen ITO (100 nm) /SnO$_2$ (2 nm) / hematite (10 nm) stack. This was done by optical simulations with a customized optical calculation algorithm (see SI) based on the transfer matrix method[44] using the optical properties of the individual layers within the stack. This code enables simulation of the spectral response of complex multilayer stacks of absorbing layers with smooth interfaces. It considers any resonant effects (if existing) and can be used with any incident illumination angle. The code is included in the supporting information.

The optical properties of the individual layers were measured by spectroscopic ellipsometry as presented in Figure S8 to Figure S12. The complex refractive indices (optical parameters) of all the layers in the stack extracted from the ellipsometry measurements are presented in Figure S13. To validate our optical simulations, we compared the calculated transmittance and reflectance spectra of our samples against the empirical spectra measured by spectrophotometry as presented for the DBR alone in Figure S14 and for the optimal full stack in the next section.

Using the optical calculation algorithm, we calculated the absorption within the hematite layer of the chosen ITO (100 nm) / SnO$_2$ (2 nm) / hematite (10 nm) stack coupled directly on top of a DBR stack comprising six identical SiO$_2$-Nb$_2$O$_5$ bilayers. Calculation of the absorption in the hematite layer using more DBR pairs did not yield significant improvement: adding another pair would increase the absorption in the hematite by only 0.2%. Figure 3 shows the dependence of the calculated absorption within the hematite layer on the SiO$_2$ and Nb$_2$O$_5$ layers thickness. The absorbed photon flux, $\varphi_{abs}$, is presented in units of photocurrent density, $J_{abs} = q \times \varphi_{abs}$, where q is the electron charge. This yields an upper limit estimation of the maximal photocurrent that may possibly be reached if every absorbed photon within the hematite layer contributes to the photocurrent, i.e., if the APCE is 100% for all photons. The simulations in Figure 3 show that a maximal photocurrent density of 3.8 mAcm$^{-2}$ may be obtained (with APCE of 100%) for a DBR stack comprising six pairs of 80 nm thick SiO$_2$ layers and 50 nm thick Nb$_2$O$_5$ layers.

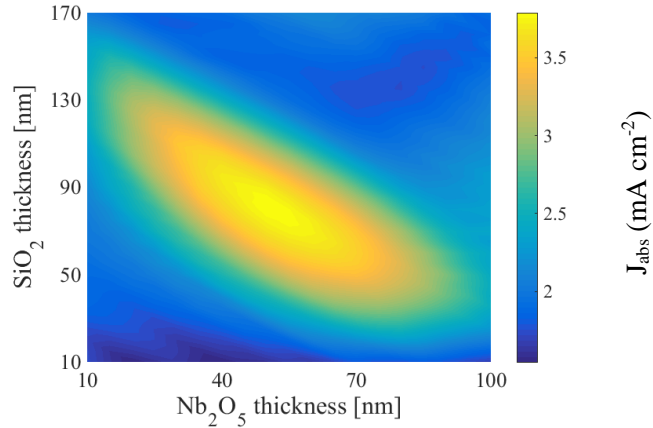

**Figure 3:** Simulated light harvesting yield, expressed as $J_{abs} = q \times \varphi_{abs}$ (see text for details), within the hematite layer of the photoanode stack as a function of the SiO$_2$ and Nb$_2$O$_5$ layer thickness of DBR stacks comprising six identical bilayer pairs coupled directly underneath ITO (100 nm) / SnO$_2$ (2 nm) / hematite (10 nm) stack.

Since many photons escape out of the photoanode by back-reflection, as discussed in the next section, further enhancement can be readily obtained by adding another (second) photoanode facing the first one in a V-shaped cell configuration, as illustrated in Figure 4. The V-shaped structure gives rise to multiple reflections between the first and second photoanodes, thereby re-trapping the escaped photons as discussed elsewhere.[22]

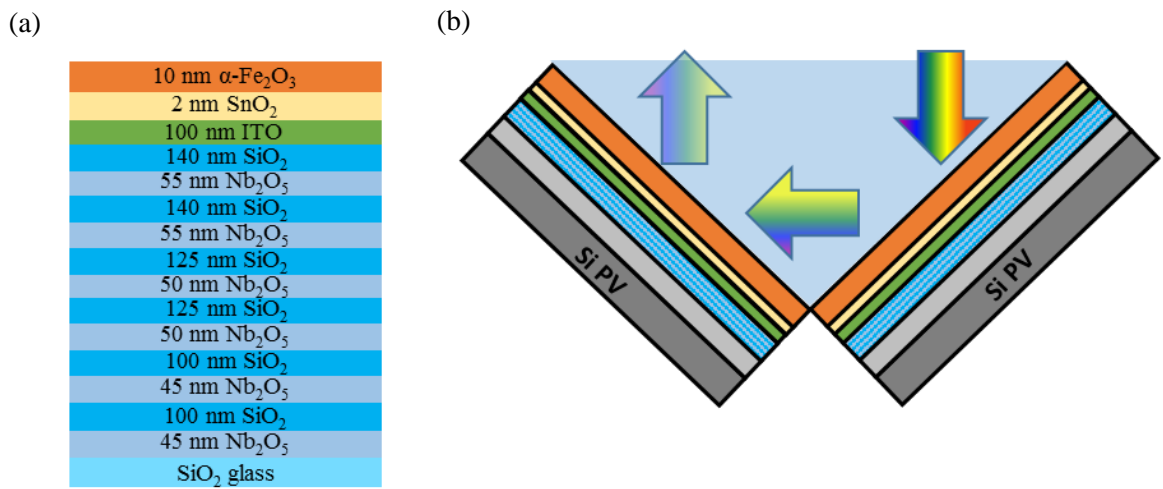

**Figure 4:** (a) Schematic representation of the target photoanode stack configuration (not to scale). (b) Schematic representation of the 90° V-shaped tandem cell configuration (not to scale).

Using a V-shaped cell configuration, the incidence angle of the illuminating light is no longer normal to the photoanode and the DBR layer thickness must be re-optimized to account for the respective incidence angles (that depend on the angle between the two plates) in order to maximize the photocurrent. Therefore, the thickness of the six DBR pairs was re-optimized for a 90º V-shaped cell. Ideally, the thickness of the ITO layer and every layer in the DBR stack should be optimized individually to achieve maximal performance. However, this optimization involves many free parameters and is prohibitively computationally intensive. In order to relax the optimization complexity, we used the following constraints: (1) we fixed the ITO thickness to the optimal thickness found without the DBR layers, and (2) we divided the DBR into three sets, each containing two identical $SiO_2$ and two identical $Nb_2O_5$ layers. Coupling of the $SiO_2$ layer thicknesses and the $Nb_2O_5$ layer thicknesses for each set reduced the fitting from 12 free parameters to 6. The structure of the optimized stack is illustrated in Figure 4. The simulated optical response of two identical optimized stacks positioned in a 90º V-shaped cell configuration illuminated in water is presented in Figure S15. Based on these calculations, the absorption in the hematite layer ($J_{abs}$) could reach to as much as 5.3 mAcm$^{-2}$, which corresponds to 44% of the photons with wavelength ranging from 350 to 590 nm, taken as the lower and upper boundaries in our optical calculations.

**Device fabrication and characterization**

Two photoanodes were fabricated according to the optimized DBR/ITO/SnO$_2$/hematite stack design. In order to reduce the onset potential, we deposited an ultrathin Fe$_{1-x}$Ni$_x$OOH ("FeNiO$_x$") co-catalyst overlayer[35,45,46] on top of the hematite layer in each stack. The co-catalyst overlayer was deposited by a photoelectrochemical deposition technique that yields a Fe:Ni ratio of ~7:1, as described elsewhere.[35] This co-catalyst overlayer was selected based on our previous work showing that ultra-thin layers of this co-catalyst are transparent and they do not alter the photoanode's optical properties.[35] Cross-sectional TEM-HAADF and TEM micrographs of the full stack are presented in Figure 5 and Figure S16, respectively, revealing smooth and well defined layers. The layer thicknesses extracted from Figure 5 and Figure S16 are presented in Table S1.

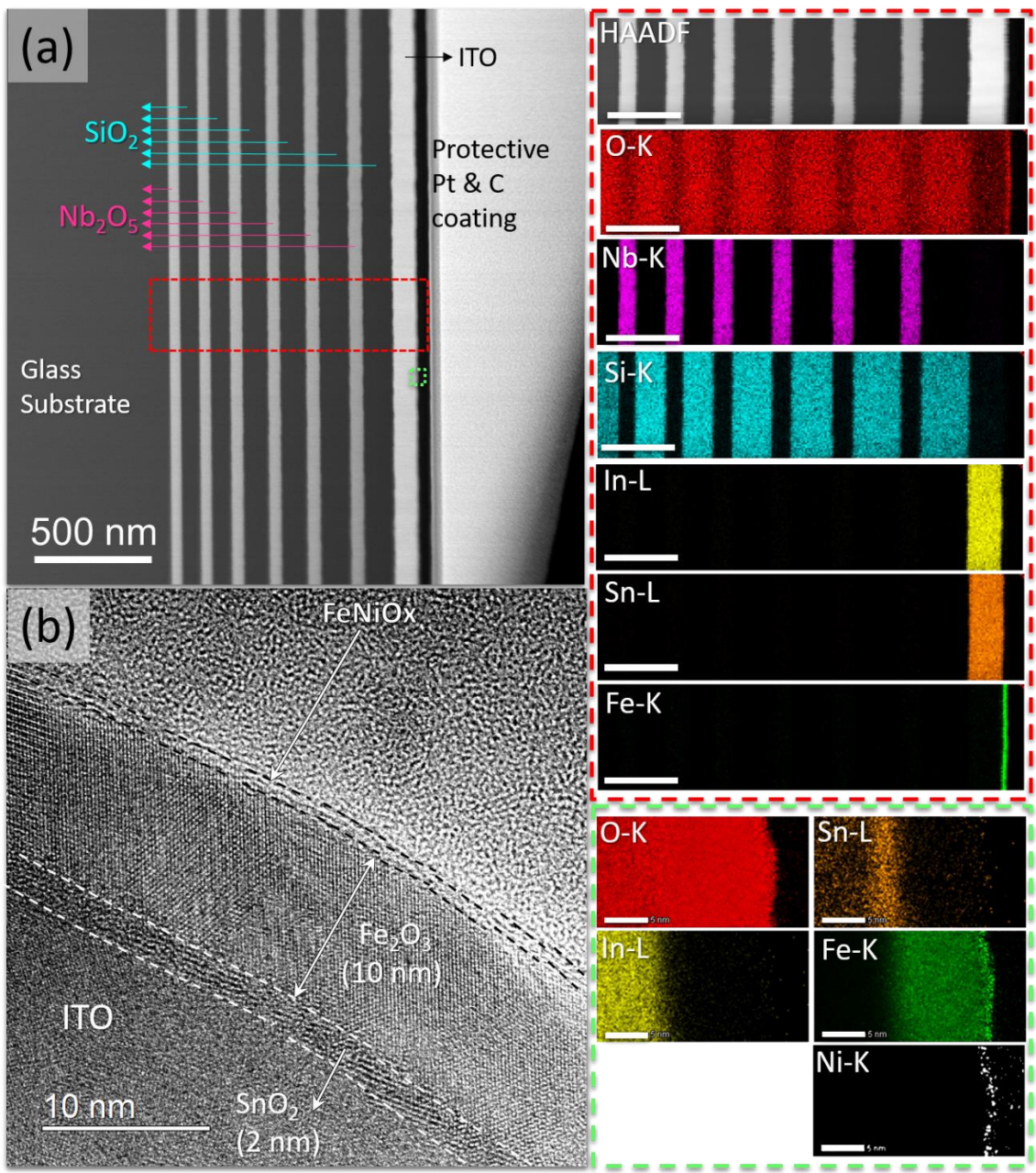

**Figure 5** (a) Cross-sectional TEM-HAADF micrograph of a full stack hematite/SnO$_2$/ITO/ DBR photoanode deposited on an eagle glass substrate. The right-hand-side panels show energy dispersive X-ray spectroscopy (EDS) elemental maps from the selected areas in (a) covering the whole stack (red, top panels) and the top layers (green, bottom panels). The scale bar in red and green panels are 200 nm and 5 nm respectively. (b) TEM image of a thin film hematite photoanode with ultrathin Fe$_{1-x}$Ni$_x$OOH ("FeNiO$_x$") co-catalyst overlayer.

**Optical properties**

Using the layer thicknesses obtained from the cross-sectional TEM measurements (Figure 5, Figure S16 and Table S1) and the optical parameters of each layer in the photoanode stack (Figure S13), obtained by spectroscopic ellipsometry of the individual layers (Figure S8 to Figure S12) we modeled the transmittance (T), reflectance (R) and absorptance (A = 1 – T – R) spectra of the photoanode stack. First, we measured the stack by spectroscopic ellipsometry and used the analysis software (WVASE32, J.A. Woollam Co.) to fit the measured $\Delta$ and $\Psi$ spectra. All the fitting parameters were set by independent measurements, except for the optical parameters of the $Nb_2O_5$ layers that were tuned to fit the measured ellipsometry spectra of the full stack. The results are shown in Figure S13. Next, the transmittance, reflectance and absorptance spectra of the stack were calculated using a customized optical calculations algorithm based on the transfer matrix method[44] (see SI). The simulated spectra are compared in Figure 6 to the empirical spectra measured by spectrophotometry. Due to small inhomogeneity in the thickness of the layers deposited by PLD (i.e, hematite and ITO), which we estimated to give rise to up to ~10% thickness variation across the entire sample, the thickness of these layers was slightly tuned (see Table S1) to obtain good agreement between the simulated and measured spectra. Figure 6(a) compares the simulated and the measured optical spectra of the full 15 layer stack (prior to the $Fe_{1-x}Ni_xOOH$ co-catalyst deposition) in air. The agreement is excellent at short wavelengths (< 900 nm), whereas at long wavelengths (> 900 nm), well above the hematite absorption edge, there is a small deviation between simulated and measured spectra. The black curves in Figure 6(a) correspond to the absorptance in the bare stack (without the $Fe_{1-x}Ni_xOOH$ co-catalyst overlayer) in air. It is noteworthy that the deviation between the simulated and measured absorptance spectra is very small for sub-bandgap photons (< 590 nm). The measured and simulated spectra of the DBR stack without the ITO, $SnO_2$ and hematite layers are shown in Figure S14. The parasitic absorptance in the DBR stack is very low, and is limited to short wavelengths (< 400 nm) where the flux of photons in the solar spectrum is relatively small. Moreover, most of the short-wavelength photons are absorbed in the hematite layer and do not reach the DBR stack below the hematite. Therefore, the parasitic absorptance in the entire stack is negligibly small, as shown in Figure S17. The black and orange curves in Figure S17 present the calculated absorptance spectra in the entire stack (overall absorptance) and in the hematite layer (hematite absorptance), calculated by our optical simulator for the bare stack (without Fe$_{1-x}$Ni$_x$OOH overlayer) illuminated from the front side in water. The ratio between the hematite and overall absorptance in water is depicted in Figure 6 (b), showing that most of the light is absorbed in the hematite layer and only a small fraction is absorbed parasitically elsewhere in the stack.

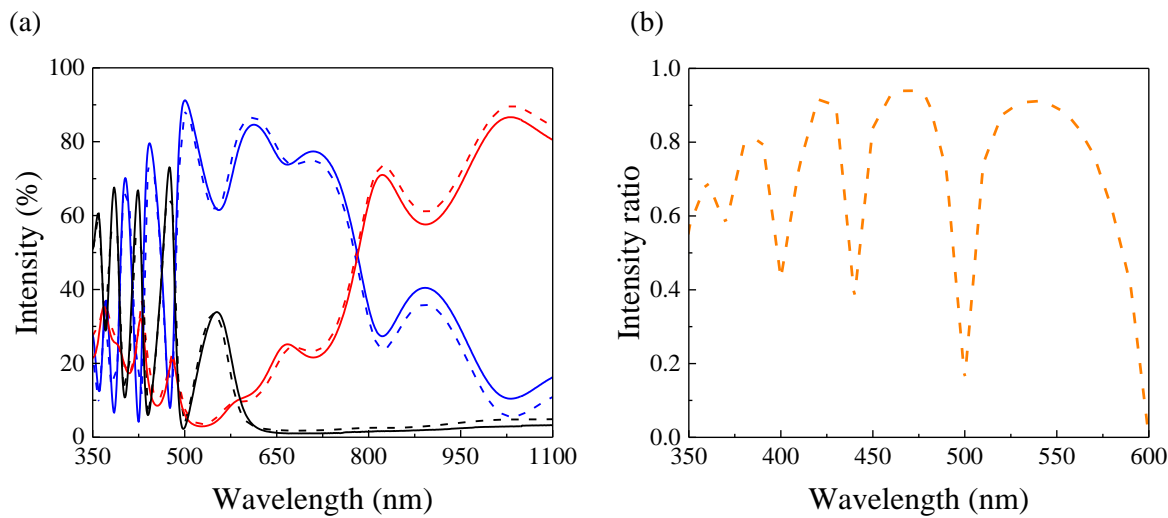

**Figure 6** (a) Measured (solid line) and simulated (dashed line) reflectance (blue), transmittance (red) and absorptance (black) spectra of the bare stack (without the Fe$_{1-x}$Ni$_x$OOH co-catalyst overlayer) in air. (b) The simulated absorption in the hematite layer divided by the absorption of the whole sample as was calculated by our algorithm for the same stack immersed in water.

By integrating the overall and hematite absorptance spectra in Figure S17 over the solar spectrum (AM1.5G) at wavelengths shorter than the hematite absorption edge (590 nm), we calculated a total absorbed photocurrent of 3.54 mAcm$^{-2}$ in the entire stack and net absorbed photocurrent of 2.95 mAcm$^{-2}$ in the hematite layer. Accordingly, 83% of the absorbed photons are absorbed in the 10 nm thick hematite layer and only 17% of them are lost for parasitic absorption elsewhere in the stack. This is similar to the ratio of hematite to overall absorption in the champion 26 nm thick hematite layer on a specular silver back-reflector, as reported elsewhere.[22] Most of the parasitic absorption in our photoanode stack occurs in the ITO layer, amounting to 0.54 mAcm$^{-2}$, which is 15% of the total absorption in the entire photoanode stack.

The red curves in Figure 6(a) show that most of the transmitted photons are above the hematite absorption edge and hence they are not suitable for photocurrent generation in hematite. However, they can be absorbed in the bottom silicon PV cell in the PEC-PV tandem cell. The blue curves show that some of the photons that can be absorbed in the hematite layer are back-reflected from the stack. These escaped photons can be re-trapped by using the V-shaped cell configuration as reported elsewhere.[22]

**Photoelectrochemical performance**

To test the photoelectrochemical performance of our photoanodes, we positioned them in a 90° V-shaped cell configuration, as illustrated in Figure 7, and measured their voltammograms in the dark and under solar simulated illumination. A small aperture (2.25 ± 0.03 mm diameter) in a blackened aluminum foil was used to define a small enough illumination area to be fully enclosed by the exposed hematite area of one of the photoanodes, whereas the other photoanode was exposed to the reflected light from the first photoanode, as illustrated in Figure 7(a).

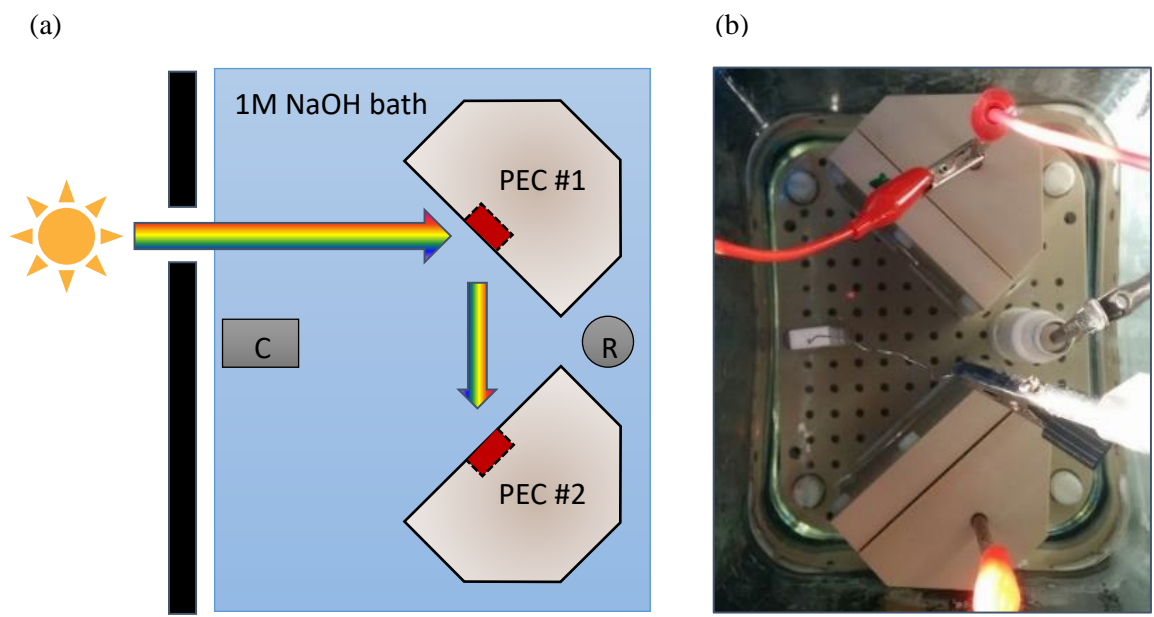

**Figure 7**: Top-view illustration (a) and photograph (b) of the experimental setup for photoelectrochemical measurements in a 90° V-shaped cell configuration. PEC #1 and PEC #2 are composed of 10 nm thick hematite, 2 nm thick $SnO_2$ under-layers and ~100 nm thick ITO transparent electrode (current collector) on identical DBRs of 6 varying bilayers pairs deposited on top of eagle glass substrates. PEC #1 is directly exposed to solar simulated light passing through a small aperture while PEC #2 is exposed only to the reflected light from PEC #1. R and C represent the reference and the counter electrodes, respectively. Front view of the the measurement setup is presented in Figure S18.

Figure 8 presents the photocurrent voltammograms measured under solar simulated illumination in alkaline solution (1M NaOH), with no sacrificial reagents, for photoanodes comprising 10 nm thick hematite layer on 2 nm thick $SnO_2$ underlayer on 100 nm thick ITO transparent electrode in the following configurations: (1) a planar photoanode without DBR and without $Fe_{1-x}Ni_xOOH$ co-catalyst overlayer (red curve); (2) a planar photoanode with DBR but without $Fe_{1-x}Ni_xOOH$ co-catalyst overlayer (green curve); (3) a planar photoanode with full stack configuration with DBR and $Fe_{1-x}Ni_xOOH$ co-catalyst overlayer (cyan curve); and (4) two photoanodes with full stack configuration with DBR (as in Figure 4) and $Fe_{1-x}Ni_xOOH$ co-catalyst overlayer (as in Figure 5) facing each other in 90° V-shaped cell configuration (as in Figure 7, blue curve). The photocurrent density was calculated by subtracting the dark current from the light current and dividing the result by the aperture area through which the cell was illuminated. We estimate the maximal measurement error using the 90° V-shaped cell configuration to be ±0.03 mAcm$^{-2}$ due to the small apreture area.

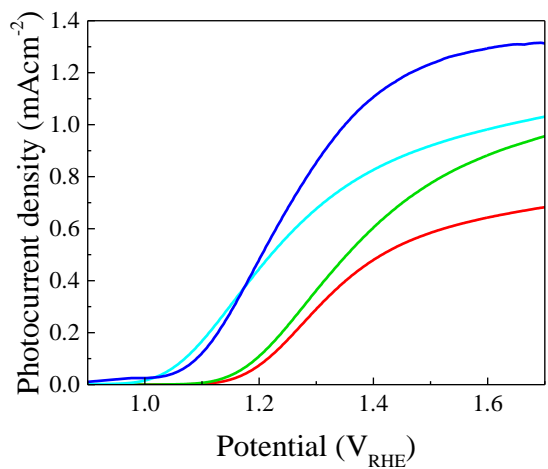

**Figure 8:** Photocurrent voltammograms of photoanodes comprised of 10 nm thick hematite, 2 nm thick $SnO_2$ underlayer and ~100 nm thick ITO transparent electrode (current collector) deposited on top of eagle glass substrates in planar configuration (red), with DBR in planar configuration (green), with DBR and $Fe_{1-x}Ni_xOOH$ co-catalyst overlayer in planar configuration (cyan), and two photoanodes with DBR and $Fe_{1-x}Ni_xOOH$ co-catalyst overlayer facing each other in the 90° V-shaped cell configuration as in Figure 7 (blue). See Figure S19 for details on the 90° V-shaped cell configuration performance.

Figure 8 shows that adding DBR alone enhances the photocurrent density in the plateau region (at 1.6 $V_{RHE}$) of the planar 10 nm thick hematite photoanode by 37%, adding the $Fe_{1-x}Ni_xOOH$ co-catalyst overlayer shifts the onset potential by 130 mV and enhances the photocurrent density in the plateau region by another 15%, and adding the 90º V-shaped cell configuration enhances the photocurrent density in the plateau region by another 48%. All in all, the photocurrent density in the plateau region was doubled by adding the DBR, co-catalyst overlayer and 90° V-shaped cell configuration. At the reversible potential of water oxidation, 1.23 $V_{RHE}$, the photocurrent density was enhanced by a factor of four compared to the control planar photoanode without DBR, co-catalyst and V-shaped cell configuration. The remarkable enhancement was achieved despite technical losses resulting from the experimental setup (shown in Figure 7), that could be significantly reduced in an optimized setup with larger photoanodes. First, the long distance between the aperture and the first illuminated photoanode, as well as the distance between the first and second photoanode, result in optical loss due to light scattering. Second, there is additional optical loss due to shadowing, mainly by the apertures in the housings of the photoanodes. Third, additional optical loss arises from imperfect alignment of the two photoanodes. These losses are estimated to add up to a total optical loss of 24% (see SI for details). This loss could be rectified in an optimally designed system with larger photoanodes, resulting in an estimated plateau photocurrent density of 1.7 mA/cm$^2$ (at 1.6 $V_{RHE}$).

**PEC-PV tandem cells**

To demonstrate overall solar water splitting, we constructed a PEC-PV tandem cell by stacking a PEC cell with a hematite photoanode (with DBR and $Fe_{1-x}Ni_xOOH$ co-catalyst overlayer, see Figure S20) and a platinum wire cathode in front of a PV mini-module with three monocrystalline Si PV cells connected in series (so as to provide a voltage high enough to complement the photovoltage generated at the PEC cell to drive both water oxidation and reduction reactions). This set up is illustrated in Figure S21. The illuminated area of the hematite photoanode was 1.47 cm$^2$, whereas the area of the Si PV cells in the PV mini-module was 1.08 cm$^2$. The mismatch in the active areas of the PEC and PV cells indicates non-ideal optical coupling that could be rectified with properly matched cells.

Likewise, the electrical coupling between the PEC and PV cells was not ideal,[2] as the crossover point of the photocurrent – voltage curves of the two cells was far away from the maximum power point of the PV mini-module, as shown in Figure 9(a). Nevertheless, despite these coupling losses, a solar-to-hydrogen (STH) conversion efficiency of 0.9% is obtained at the crossover point of the two cells. Figure 9(b) presents the short-circuit current obtained with this tandem cell over 100 min operation. The short-circuit current was larger than 1 mA, which corresponds to a current density of 0.7 mAcm$^{-2}$ and an STH of 0.8% considering the illuminated area of the hematite photoanode in the PEC cell. The slow current drift resulted from bubble accumulation, as can be seen by the sudden jumps in Figure 9(b) when large bubbles were released.

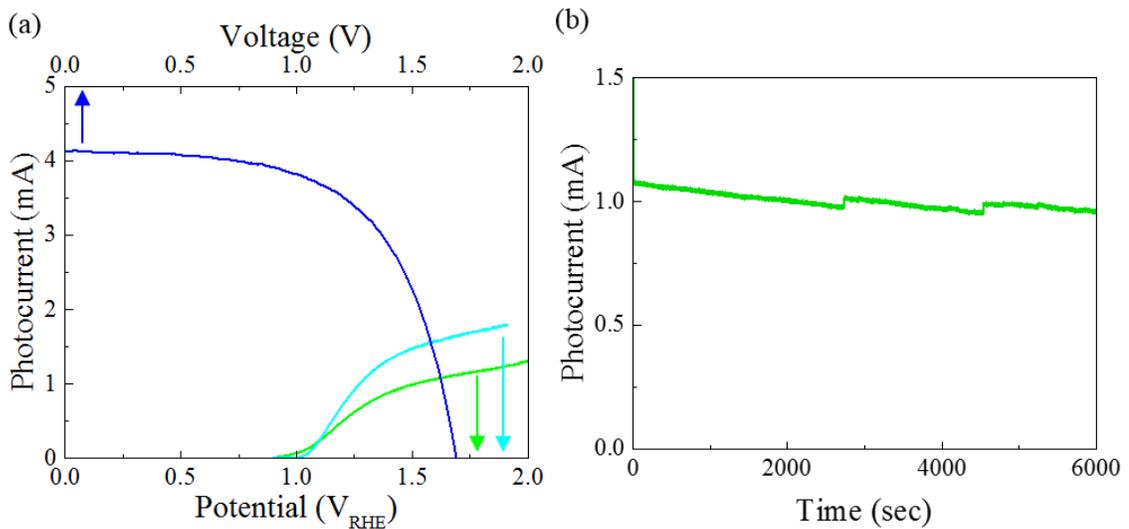

**Figure 9:** (a) Photocurrent voltamogram of a non-optimized hematite photoanode with DBR and Fe1-xNixOOH co-catalist overlayer (green) and photocurrent – voltage curve of a silicon PV mini-module behind the same photoanode (blue). The photocurrent density of the photoanode stack from Figure 8 (without the V-shape cell, cyan curve) multiplied by the aperture area of this set-up is shown by the cyan curve. (b) Stability test of self-asssisted solar water splitting performed along 6000 seconds measured with a hematite photoanode fitted with a non-optimized DBR.

**Discussion on further potential enhancements**

The hematite photoanode used for demonstrating overall solar water splitting in Figure 9 was somewhat different than the optimized stack (see Figure 8 for details on the optimized stack performance). Using the optimized stack, and assuming same PV performance as the measured results, will increase the STH efficiency to 1.3% (see cyan curve in Figure 9(a)). Additional photocurrent enhancement may be obtained by using narrow angle (< 90°) V-shaped cell configurations which result in multiple reflections. Assuming 100% IPCE, the maximal photocurrent density that may be obtained using a 90° V-shaped cell configuration is 5.3 mAcm$^{-2}$. This number represents the calculated net absorption within the 10 nm hematite layer of the total stack. Using smaller angles between the two photoanode stacks at the plates of the V-shaped cell will result in more reflections between them, thereby enhancing the photocurrent. Figure 10 presents the expected net absorbed photocurrent ($J_{abs}$) within the hematite layer for different angles of the V-shaped structures, as simulated using our optical calculation algorithm for the previously described full stack (Figure 4). According to these calculations, using a 30° V-shaped structure may result in $J_{abs}$ of 6.8 mAcm$^{-2}$, which corresponds to ~57% of all the photons that can be absorbed in hematite for standard solar illumination in this spectral range (350 – 590 nm). Calculations of the maximum amount of light that can be absorbed in same thickness (10 nm) hematite layer on a planar transparent substrate result in $J_{abs}$ of 2.0 mAcm$^{-2}$. Therefore, our DBR stack combined with a 30º V-shaped structure can more than triple the light harvesting yield within a 10 nm thick hematite layer. It is noted that the absorption for planar configuration with this stack is lower than the absorption previously calculated with optimized six DBR layers because this DBR stack was tailored to maximize absorption for 90° V-shape cell configuration. Ergo, conceptually, designated optimization for the 30° V-shape structure is expected to achieve even higher $J_{abs}$ than 6.8 mAcm$^{-2}$.

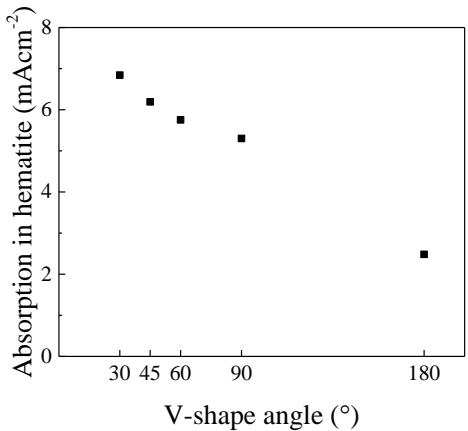

**Figure 10:** Calculated net absorbed photocurrent ($J_{abs}$) in a 10 nm thick hematite layer within the optimized stack configuration for different V-shape angles. A 180° V-shape angle represents planar configuration (i.e. no V-shape). An example for the multiple reflected light paths for the 30º V-shape structure is presented in Figure S22.

Improving the productive absorption, and hence, the APCE of the hematite photoanode is critical towards improved device performance. Recent work introduced the spatial collection efficiency of photogenerated charge carriers in photoelectrochemical cells[47] considering the photogeneration efficiency, defined as the probability of absorbed photons to generate mobile charge carriers at a given wavelength. Due to the existence of both localized excitations which do not produce mobile charge carriers and delocalized excitations which do, the photogeneration efficiency changes dramatically with illumination wavelength and significantly affects the photoelectrochemical performance of hematite photoanodes.[48] Tailoring new DBRs to account for productive vs. nonproductive optical transitions could potentially further improve the photoanode performance.

**Conclusion**

Unassisted solar water splitting was demonstrated with a PEC-PV tandem cell comprising an ultrathin film hematite photoanode stacked in front of a silicon PV mini-module using a wavelength-selective dielectric mirror (a.k.a. distributed Bragg reflector, DBR) designed for optimal optical coupling between the PEC and PV mini-module. The DBR can be tailored to maximize light harvesting within the ultrathin hematite film using strong interference, resulting in a more than threefold enhancement in absorbed photocurrent density as compared to films deposited on standard transparent current collectors. Unlike metallic back-reflectors, the DBR can be tailored to transmit the part of the solar spectrum that is not absorbed in the front PEC cell to reach the silicon PV mini-module behind it, thereby enabling stacked PEC-PV tandem cell configuration while employing the benefits of strong interference in ultrathin films. This approach can be employed for enhancing the light harvesting efficiency of any ultrathin absorbing layer with smooth interfaces in a tandem configuration.

**Methods**

A 1% cation Sn-doped $Fe_2O_3$ target was prepared using high purity powders of $Fe_2O_3$ and $SnO_2$ (99.9% and 99.998% respectively). The powders were mixed by mortar and pestle, moistened with isopropanol, mixed again using low energy milling machine, dried, mixed with a binder, pressed into molds, cold isostatic pressed and finally sintered at 1250°C for 4 h (in air). The target composition was examined by energy dispersive X-ray spectroscopy (EDS, Zeiss Ultra-Plus HRSEM), confirming the expected composition with 1.03±0.09 cation% of Sn. The analysis was carried out at an acceleration voltage of 10 kV, working distance of 8.3 mm, and 180 s livetime. Iron K and tin L lines were used to analyze the data by oxygen stoichiometry, taking the oxidation states of iron and tin as +3 and +4, respectively.

The $SiO_2$ and $Nb_2O_5$ layers in the DBR stacks were deposited from commercially supplied targets (99.99% purity, AJA) using an AJA ATC 2200 RF sputtering system. The sputter deposition was carried out at 300°C in 50sccm:5sccm $Ar:O_2$ flow at a pressure of 3 mTorr. The $SiO_2$ and $Nb_2O_5$ layers were deposited by sputtering instead of PLD because the laser beam (KrF, $\lambda$=248 nm) in our PLD system is not suited for $SiO_2$ ablation. In addition to the Sn-doped $Fe_2O_3$ target, undoped $SnO_2$ (purity 99.9%) and ITO (purity 99.99%, $In_2O_3/SnO_2$ ratio of 90/10) targets were purchased from Kurt J. Lesker and ACI Alloys, respectively. The hematite layers, $SnO_2$ underlayers and ITO current collectors were deposited from these targets by pulsed laser deposition (PLD) using a turn-key PLD workstation (PVD Products, USA) with a KrF pulsed excimer laser beam ($\lambda$=248 nm). All the layers were deposited at a set-point temperature of 300°C and in oxygen pressures of 3 mTorr (ITO films), 100 mTorr ($SnO_2$ films) and 25 mTorr (hematite films). Following the deposition, the samples were annealed at 500°C for two hours in air. The $Fe_{1-x}Ni_xOOH$ co-catalyst overlayer was deposited on the hematite photoanodes by photoelectrochemical deposition, as described elsewhere.[35,45] The composition of the solution was 5mM $Fe_2(SO_4)_3$, 16mM nickel(II) acetate and 0.1M sodium sulfate. The deposition was performed under constant current conditions and the illumination was provided by Mightex white LED system (6500K). The thickness of the $Fe_{1-x}Ni_xOOH$ overlayer was ~1.5 nm and the Fe:Ni ratio was ~7:1, as reported elsewhere.[35]

The layer thicknesses of the full stack were obtained directly by cross-sectional transmission electron microscopy (TEM). The cross-sectional TEM sample was prepared by dual-beam focused ion beam (FEI Helios NanoLab G3 FIB). The sample was externally coated (Q150T Quorum Technologies, carbon coater) with 50 nm thick carbon to protect the surface of the hematite photoanode from the platinum deposition and ion milling in the standard lift out process used in the focused ion beam (FIB). The prepared TEM lamella was further gentle milled (Technoorg Linda Ltd) and plasma cleaned (model 1020 Fischione instruments) before its placement in the TEM column. The high-resolution TEM, high angle annular dark field (HAADF-STEM) imaging and elemental mapping was performed by using monochromated and double aberration-corrected (CEOS) Titan Cubed Themis G2 300 (FEI / Thermo Fisher) microscope operated at 300 kV. The surface roughness was examined by atomic force microscopy (AFM, MFP-3D Infinity AFM, Asylum Research, Oxford instruments). The ITO sheet resistance was measured by 4 point probe technique (Signatone, using Keithley current source and multimeter). Optical properties of the stacks were measured by spectrophotometery using a Cary 5000 series UV-Vis-NIR spectrophotometer (Agilent Technologies). The total absorption was obtained from transmission and reflection measurements using a Universal Measurement Accessory (UMA). The optical parameters (complex refractive index) of all the layers were obtained by spectroscopic ellipsometry (VASE Ellipsometer, J.A. Woollam) using samples with single layers deposited on eagle glass substrates, except for the $SnO_2$ layer that was deposited on silver and the $SiO_2$ layer whose optical constants were taken from the library of the ellipsometry analysis software (WVASE32, J.A. Woollam Co.). The photocurrent was measured by linear sweep voltammetry in 1M NaOH solution (in deionized water, without any sacrificial reagents) under solar simulated (AM 1.5G) illumination using a Sun 3000 class AAA solar simulator (ABET Technologies). 3 electrode voltammetry measurements were carried out using a potentiostat (CompacStat; Ivium Technologies B. V., Eindhoven, The Netherlands) with an Hg/HgO in 1M NaOH reference electrode (ALS Co., Ltd) and a Pt wire counter electrode (ALS Co., Ltd). Photoelectrochemical measurements of planar photoanodes were careered out using "cappuccino cells" as reported elsewhere.[43] Unassisted photoelectrochemical water splitting tests were carried out with a Si PV mini-module (KXOB22-04X3L, IXOLAR™ SolarBIT) behind the PEC cell. This 22 mm × 7 mm × 1.8 mm mini-module is comprised of three cells connected in series yielding as open circuit voltage ($V_{oc}$) of 1.89 V and short circuit current ($I_{sc}$) of 15 mA. The solar-to-hydrogen (STH) efficiency was calculated according to equation (1) in reference [9].

**Supporting Information**

Supplementary Information includes additional explanations, figures and the optical modeling code that was originally written by Dr. Hen Dotan and Dr. Ofer Kfir and later modified by Yifat Piekner.

**Acknowledgments**

This research has received funding from the European Research Council under the European Union's Seventh Framework programme (FP/2007-2013) / ERC (grant agreement n. 617516). Y.P. acknowledges support by a scholarship from the Ministry of National Infrastructures, Energy and Water Resources of Israel (2016), and by the Levi Eshkol scholarship from the Ministry of Science and Technology of Israel (2017). D.A.G. acknowledges support by Marie-Sklodowska-Curie Individual Fellowship n. 659491. The experiments were performed using central facilities at the Technion's Photovoltaics Research Laboratory supported by the Adelis Foundation, the Grand Technion Energy Program (GTEP), and the Russel Berrie Nanotechnology Institute (RBNI). The authors thank Dr. Guy Ankonina for deposition of the DBR layers using the sputter deposition system at the Technion's Micro-Nano Fabrication Unit (MNFU), and Dr. Cecile Saguy for taking the AFM image. The authors acknowledge technical assistance by the team of the Technion's Electron Microscopy Center (MIKA).

**Author contributions**

Y.P. deposited the PLD layers, simulated the optical performance, conducted all measurements and analyzed the ellipsometery measurements. H.D. and A.R. conceived the idea. A.R., H.D and D.A.G. advised and supervised the project. A.T. deposited the $Fe_{1-x}Ni_xOOH$ co-catalyst overlayer. K.D.M and Y.P. took the TEM images. K.D.M performed the EDS analysis. H.D. and O.K. wrote the original optical modeling code that was later modified by Y.P. H.D. conceived and constructed the 90° V-shaped configuration set-up. Y.P. constructed the PEC-PV tandem configuration set-up. Y.P. wrote the original draft of the manuscript. A.R., D.A.G and Y.P. reviewed and edited the manuscript.

**Conflict of Interest**

The authors declare no conflict of interest.

**TOC graphic**

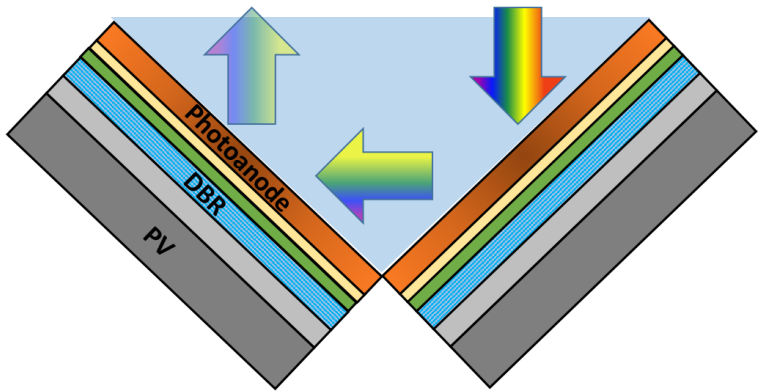